\documentclass[12pt]{iopart}
\pdfoutput=1
\usepackage{iopams}
\usepackage{graphicx}
\usepackage{mathrsfs}
\usepackage{gensymb}
\usepackage{hyperref}

\begin{document}
\title[Delta(1232) contribution to real radiative corrections for elastic
$ep$-scattering]{Delta(1232) contribution to real photon radiative corrections
for elastic electron-proton scattering}

\author{R E Gerasimov$^{1,2}$ and V S Fadin$^{1,2}$}
\address{$^1$Budker Institute of Nuclear Physics, Novosibirsk, Russia}
\address{$^2$Novosibirsk State University, Russia}
\eads{\mailto{r.e.gerasimov@inp.nsk.su}, \mailto{fadin@inp.nsk.su}}

\begin{abstract}
    Here we consider a contribution of Delta(1232) resonance to real photon radiative
    corrections for elastic $ep$-scattering. The effect is found to be small for
    past experiments to study unpolarized cross section as well as for the
    recent VEPP-3 experiment to investigate two-photon exchange effects by
    precision measurement of $e^\pm p$-scattering cross sections ratio.
\end{abstract}
\vspace{2pc}


\maketitle

\section{Introduction}

The electromagnetic form factors of the proton ($G_{\rm E,M}$) contain information
about its internal structure. The Rosenbluth separation method
\cite{PhysRev.79.615} have been used since 1950s to extract the form factors
from the unpolarized electron-proton scattering cross section \cite{PhysRev.98.217,
PhysRev.102.851,RevModPhys.28.214,PhysRevLett.5.261,PhysRevLett.5.263,
RevModPhys.35.335,PhysRev.142.922,PhysRevLett.20.292}.
It became possible to extract the ratio $G_{\rm E}/G_{\rm M}$ with the
polarization transfer method \cite{Akhiezer:1974em} since 2000s, and the results
obtained by the two methods unexpectedly contradict each other
\cite{PhysRevLett.84.1398, PhysRevC.64.038202, PhysRevLett.88.092301,
PhysRevC.68.034325, PhysRevC.71.055202, PhysRevLett.104.242301}.

It is suggested that a more accurate account of the two-photon exchange (TPE)
effects in the experiments with unpolarized particles can reduce the discrepancy
\cite{PhysRevLett.91.142303}. Theoretical investigations of TPE effects have
been performed using various models (see reviews \cite{ISI:000251313900007,
Arrington2011782} and references therein). For example, in the hadronic model
the TPE amplitude can be approximated by successive consideration of the virtual
proton, $\Delta$(1232) and higher resonances in the proton intermediate state
\cite{PhysRevLett.91.142304, PhysRevLett.95.172503, PhysRevC.72.034612,
Zhou:2014xka}.

From the experimental point of view the TPE effects can be studied in comparison
of elastic electron-proton and positron-proton scattering cross sections. There
are three new experiments \cite{PhysRevLett.114.062003, Milner20141,
PhysRevLett.114.062005} aiming to precise measure the cross section ratio 
$R = \sigma(e^+p) / \sigma(e^-p)$. In the leading order of the electromagnetic
coupling constant we have
\begin{equation}
    R = 1 - 2 \delta_{2\gamma} - 2 \delta_{{\rm brem,odd}},
    \label{eq:Rdecomposition}
\end{equation}
where the virtual radiative correction~$\delta_{2\gamma}$ comes from the
interference of the TPE amplitude with the one-photon exchange (Born) amplitude;
the C-odd real radiative correction~$\delta_{{\rm brem,odd}}$ originates from
the interference of the electron and proton bremsstrahlung amplitudes. The both
corrections have infrared divergences which cancel in their sum. The Mo-Tsai
convention \cite{RevModPhys.41.205} is commonly used to regularize and
pick out the soft photon terms:
\begin{equation}
    \delta_{2\gamma} =
    \delta_{2\gamma}^{{\rm soft}}
    + \delta_{2\gamma}^{{\rm hard}},
   \qquad 
\delta_{{\rm brem,odd}} =
    \delta_{{\rm brem,odd}}^{{\rm soft}}
    + \delta_{{\rm brem,odd}}^{{\rm hard}}.
\end{equation}

To extract the TPE effects contribution $\delta_{2\gamma}^{{\rm hard}}$ from $R$
it is necessary to exclude the ``hard'' part of real radiative corrections
$\delta_{{\rm brem,odd}}^{{\rm hard}}$.  This correction strongly depends on
particular experimental conditions. In this paper we will primarily address to
the experiment at the VEPP-3 storage ring \cite{PhysRevLett.114.062005}. The
ESEPP event generator \cite{0954-3899-41-11-115001} was used to calculate
$\delta_{{\rm brem,odd}}^{{\rm hard}}$ for the VEPP-3 experiment. It takes into
account virtual proton intermediate state in proton bremsstrahlung.  Using the
hadronic model one has to consider resonances in the intermediate state.
Their contributions do not have infrared divergences since the spectrum of
bremsstrahlung photons in this case is different from the infrared $\rmd \omega
/ \omega$, because the resonances have masses distinct from the mass of the
proton, that prevents the appearance of $\omega$ in the denominator.  We can
expect that $\Delta(1232)$ will give the leading contribution since it is the
lowest resonance as well as it has considerable branching for the decay $\Delta
\to p \gamma$. As we will show in the following a naive estimate gives
significant contribution to the radiative corrections, and only a more accurate
calculation ensures us that this correction is actually small.

\section{Transition vertexes and form factors}

Let us consider the process $\gamma(q)\, p(p) \to \Delta(p_\Delta)$. We will use
the following prescription for the transition matrix element:
\begin{equation}
    \rmi \mathcal{M}_{\gamma p \to \Delta} = 
        \rmi Ze\, J_{p\to \Delta}^\nu(p, p_\Delta)\ \epsilon_\nu(q),
    \label{eq:MgpD}
\end{equation}
where the transition current
\begin{equation}
    J_{p\to\Delta}^\nu(p, p_\Delta) = 
        \bar{U}_\beta(p_\Delta)\,
        \Gamma_{\gamma p \to \Delta}^{\nu\beta}(p_\Delta,q)\,
        U(p),
    \label{eq:currentDefinition}
\end{equation}
the electron charge $e = - |e|$; $Z = +1$ is preserved as a matter
of traditional notation for radiative corrections to distinguish C-odd and
C-even terms; $\epsilon_\nu(q)$ is the photon polarization vector, $U(p)$ is the
proton bispinor, $\Delta$ is described with the help of the spin-$3/2$ wave
function $U_\beta(p)$.

The electromagnetic current is hermitian. From (\ref{eq:currentDefinition})
one can derive the following relation between the transition vertexes in the
direct $(\gamma p \to \Delta)$ and inversed  $(\Delta \to \gamma p)$ processes
as it was emphasized in \cite{Zhou:2014xka}:
\begin{equation}
    \Gamma_{\Delta \to \gamma p}^{\nu\beta}(p_\Delta,q) =
    \gamma^0\left(\Gamma_{\gamma p \to \Delta}^{\nu\beta}(p_\Delta,q)\right)^\dagger\gamma^0,
    \label{eq:vertexRelation}
\end{equation}
where in both sides $p_\Delta$ stands for the $\Delta$ momentum, $q$ is the
photon momentum, and $p = p_\Delta-q$ is the proton momentum.

Zhou and Yang \cite{Zhou:2014xka} make use of the following parameterization:
\begin{eqnarray}
    \fl \Gamma_{\gamma p \to \Delta}^{(ZY),\,\nu \beta}(p_\Delta,q)
    =& -\sqrt{\frac{2}{3}}
    \frac{1}{2M_\Delta^2}\ \gamma^5 \Bigl\{ 
        G_1(q^2) \left[
            g^{\nu \beta}\hat{q}\hat{p}_\Delta - p_\Delta^\nu\hat{q}\gamma^\beta -
            \gamma^\beta\gamma^\nu(p_\Delta\cdot q) + \hat{p}_\Delta\gamma^\nu q^\beta
        \right] \nonumber\\
        &+ G_2(q^2) \left[ p_\Delta^\nu q^\beta - g^{\nu\beta}(p_\Delta\cdot q) \right] \nonumber\\
        &- \frac{G_3(q^2)}{M_\Delta} \left[ 
            q^2 \left( p_\Delta^\nu \gamma^\beta - g^{\nu\beta} \hat{p}_\Delta \right) 
            + q^\nu \left( q^\beta \hat{p}_\Delta - \gamma^\beta(p_\Delta\cdot q) \right)
        \right]
    \Bigr\},
    \label{eq:ZY}
\end{eqnarray}
where $\gamma^5 = \rmi\gamma^0 \gamma^1 \gamma^2 \gamma^3$, and 
$M_\Delta$ is the $\Delta(1232)$ mass. The form factors $G_i$
depend only on $q^2$, so they are real functions in the region $q^2 < 0$, where
we do not have any discontinuities.
In the following to provide numerical results we apply the model
from~\cite{Zhou:2014xka}, which defines the form factors
\begin{equation}
G_i(q^2) = g_iF_\Delta^{(1)}(q^2),\qquad i = 1,2,3,
\end{equation}
by the set of parameters $\{g_1,\ g_2,\ g_3\} = \{6.59,\
9.08,\ 7.12\}$ (the values of the form factors at $q^2 = 0$) and the
$q^2$-dependant factors\
\begin{equation}
\eqalign{
& F_\Delta^{(1)}(q^2) = F_\Delta^{(2)}(q^2) = 
\left(\frac{-\Lambda_1^2}{q^2 - \Lambda_1^2}\right)^2
\frac{-\Lambda_3^2}{q^2 - \Lambda_3^2},\\ 
& F_\Delta^{(3)}(q^2) = 
\left(\frac{-\Lambda_1^2}{q^2 - \Lambda_1^2}\right)^2
\frac{-\Lambda_3^2}{q^2 - \Lambda_3^2}\left[
a\, \frac{-\Lambda_2^2}{q^2 - \Lambda_2^2} +
(1-a)\, \frac{-\Lambda_4^2}{q^2 - \Lambda_4^2}
\right],}
\end{equation}
with $\Lambda_1 = 0.84\ {\rm GeV}$, $\Lambda_2 = 2\ {\rm GeV}$, $\Lambda_3 =
\sqrt{2}\ {\rm GeV}$, $\Lambda_4 = 0.2\ {\rm GeV}$, $a = -0.3$.

There is a more commonly used parametrization by Jones and Scadron
\cite{Jones:1972ky} in terms of the magnetic $G_{\rm M}^*(q^2)$, electric
$G_{\rm E}^*(q^2)$ and Coulomb $G_{\rm C}^*(q^2)$ form factors:
\begin{eqnarray}
\fl \Gamma_{\gamma p \to \Delta}^{(JS),\,\nu \beta}(p_\Delta,q)=&
	-\rmi\sqrt{\frac{2}{3}}\,
	\frac{3(M_\Delta + M_p)}{2M_p \left[(M_\Delta + M_p)^2 - q^2\right]}
    \Biggl\{G_{\rm M}^*(q^2)\ \epsilon^{\nu\beta\rho\sigma} (p_\Delta)_\rho
q_\sigma\nonumber\\
&+ G_{\rm E}^*(q^2)\left[
\frac{4\,\epsilon^{\nu\tau\rho\sigma} (p_\Delta)_\rho q_\sigma\, g_{\tau\tau'}\,
	\epsilon^{\beta\tau'\lambda\kappa} (p_\Delta)_\lambda q_\kappa}
    {(M_\Delta - M_p)^2 - q^2}\,
    (\rmi\gamma^5)-\epsilon^{\nu\beta\rho\sigma} (p_\Delta)_\rho q_\sigma  \right]\nonumber\\
    &+ G_{\rm C}^*(q^2)\, \frac{2\left(q^2\, p_\Delta^\nu - (q \cdot p_\Delta)\,q^\nu
	\right)q^\beta}{(M_\Delta - M_p)^2 - q^2} \,(\rmi\gamma^5)
\Biggr\},
\label{eq:JS}
\end{eqnarray}
where $\epsilon^{0123} = +1$, and $M_p$ is the proton mass.

Considering the matrix element $\mathcal{M}_{\gamma p \to \Delta}$
for the definite helicities of the particles we can find the
relations between the two set of form factors:
\begin{equation}
\eqalign{
\fl G_{\rm M}^*(q^2) = \frac{M_p}{3(M_\Delta + M_p)}
\Biggl[
\frac{(M_\Delta + M_p)^2 - q^2}{M_\Delta^2} G_1(q^2)\\
-
\frac{M_\Delta^2-M_p^2 + q^2}{2M_\Delta^2}(G_1(q^2)-G_2(q^2))-
\frac{-q^2}{M_\Delta^2} G_3(q^2)
\Biggr],\\
\fl G_{\rm E}^*(q^2) = \frac{M_p}{3(M_\Delta + M_p)}\left[
-\frac{M_\Delta^2-M_p^2 + q^2}{2M_\Delta^2}(G_1(q^2)-G_2(q^2))-
\frac{-q^2}{M_\Delta^2} G_3(q^2)
\right],\\
\fl G_{\rm C}^*(q^2) = \frac{2M_p}{3(M_\Delta + M_p)} \left[
-(G_1(q^2)-G_2(q^2))+
\frac{(M_\Delta^2-M_p^2 + q^2)}{ 2M_\Delta^2 }G_3(q^2)
\right].
}
\label{eq:JS_ZY_relation}
\end{equation}
These formulas for $q^2 = 0$ can be found in \cite{Zhou:2014xka}. To
check them for $q^2 \ne 0$ it is possible to combine expressions from
\cite{Zhou:2014xka} and from the review \cite{Pascalutsa2007125}.

\section{A rough estimate for Delta(1232) contribution to real radiative
correction}

\begin{figure}
	\centering
	\includegraphics{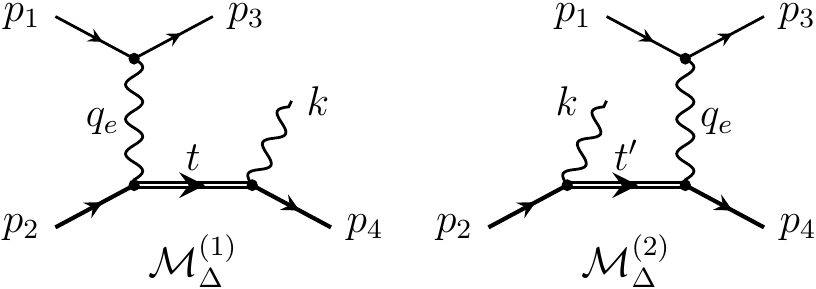}
	\caption{Feynman diagrams for the proton bremsstrahlung with $\Delta$ in the
		intermediate state}
	\label{fig:M_Delta_diagrams}
\end{figure}

Using the hadronic model we have to consider two Feynman diagrams presented
on the Figure~\ref{fig:M_Delta_diagrams}. To find relevant contribution to
radiative corrections we must calculate the square of absolute value of their sum
and the interference of these amplitudes with the amplitudes of electron and
proton bremsstrahlung. Then we have to integrate the result over the
final particles phase space taking into account the particular experimental
conditions. Divided by the elastic process cross section it yields the
contribution $\delta_{\Delta}$ to real radiative corrections for
electron-proton scattering. We will implement this procedure in the next
section. But one can note that the first amplitude on the
Figure~\ref{fig:M_Delta_diagrams} has ``resonant'' behavior: the virtual photon
energy transfer makes the intermediate $\Delta$ to be closer to the
real particle pole position so the square of this amplitude might be dominant and
give a reasonable approximation. It is worth to note that both amplitudes are
gauge invariant separately due to the interaction vertex structure, that is the
reason why they can be treated independently. Taking this into account a very rough
estimate of $\delta_{\Delta}$ can be obtained if we consider the bremsstrahlung
as two successive processes $ep\to e\Delta$ and $\Delta\to p \gamma$ and assume
that all photons from the decay contribute to real radiative corrections:
\begin{equation}\label{eq:rough_estimate}
\delta_\Delta \simeq
\frac{{\rmd\sigma'}/{\rmd\Omega}}{{\rmd\sigma}/{\rmd\Omega}} \ 
\frac{\Gamma_{\Delta\to p\gamma}}{\Gamma_\Delta},
\end{equation}
where $\rmd\sigma/\rmd\Omega$ is the differential cross section for elastic
 process $ep\to ep$ with respect to the electron scattering angle $\rmd \Omega$;
$\rmd\sigma'/\rmd\Omega$ is the differential cross section for the process $ep \to
e\Delta$ with the same electron scattering angle;
${\Gamma_{\Delta\to p\gamma}}$ and ${\Gamma_\Delta}$ are partial and
full widths of $\Delta(1232)$, their ratio defines probability for the decay to
be electromagnetic.

To use the estimate \eref{eq:rough_estimate} we need the cross section ratio
for the elastic scattering $e(p_1)\,p(p_2) \to e(p_3)\,p(p_4)$ and the process
$e(p_1)\,p(p_2) \to e(p_3')\,\Delta(p_4')$. In this section the quantities
without primes refer to $ep\to ep$, and the quantities with primes to $ep \to
e\Delta$.
The initial state is the same for the both processes: the electron and
proton 4-momenta are $p_1 = \{\varepsilon_1, \bi{p}_1\}$ and $p_2 = \{M_p,
0\}$ correspondingly.
The final states are different: electron and proton($\Delta$) 4-momenta are $p_3{}^({}'{}^) =
\{\varepsilon_3{}^({}'{}^), \bi{p}_3{}^({}'{}^)\}$  and $p_4{}^({}'{}^) =
\{\varepsilon_4{}^({}'{}^), \bi{p}_4{}^({}'{}^)\}$. The momentum transfers
are $q{}^({}'{}^)=p_1 - p_3{}^({}'{}^)=p_4{}^({}'{}^)-p_2$.
In addition to the proton $M_p$, and the $\Delta(1232)$ mass $M_\Delta$
we will use the electron mass $m$ (in most cases we consider
ultrarelativistic electrons $\varepsilon_{1},\varepsilon_{3}{}^({}'{}^) \gg m$
and $(q{}^({}'{}^))^2\gg m^2$).

The matrix elements have the following form:
\begin{eqnarray}
\rmi \mathcal{M} = -\frac{\rmi Ze^2}{q^2}\ j_{\nu}(p_1, p_3)\,J_p^{\nu}(p_2, p_4),\\
\rmi\mathcal{M}' = -\frac{\rmi Ze^2}{q'^2}\ j_{\nu}(p_1, p_3')\,J_{p\to\Delta}^{\nu}(p_2, p_4'),
\end{eqnarray}
where the electron current
\begin{equation}
j^\nu(p_1, p_3) = \bar{u}(p_3)\,\gamma^\nu\,u(p_1),
\end{equation}
the proton current
\begin{equation}
J_p^\nu(p_2, p_4) = \bar{U}(p_4)\,\Gamma_{\gamma p \to p}^\nu(q)\,U(p_2),
\end{equation}
and the transition current $J_{p\to\Delta}$ is defined by
(\ref{eq:currentDefinition}).

The proton electromagnetic vertex is parametrized with the help of two form
factors $F_{1,2}(q^2)$:\
\begin{equation}
\Gamma_{\gamma p \to p}^\nu(q) =
F_1(q^2)\,{\gamma^\nu} -
F_2(q^2)
\frac{[\gamma^\nu,\hat{q}]}{4M_{p}}.
\end{equation}
The proton electric $G_{\rm E}$ and magnetic $G_{\rm M}$ form factors can be expressed in terms of
$F_{1,2}$ as follows:\
\begin{equation}
    G_{\rm M}(q^2)= F_1(q^2) + F_2(q^2),\qquad
    G_{\rm E}(q^2) =  F_1(q^2) + \frac{q^2}{4M_p^2}F_2(q^2).
\end{equation}

The differential cross sections for unpolarized particles
in the case of ultrarelativistic electrons for the same electron
scattering angle $\theta$ (azimuthal symmetry leads to $\rmd \Omega = 2\pi\, \rmd \cos \theta$)
are
\begin{equation}
\frac{\rmd\sigma{}^({}'{}^)}{\rmd\Omega} =
\frac{1}{(4\pi)^2}\frac{1}{4M_p^2\eta}\frac{\varepsilon_3{}^({}'{}^)}{\varepsilon_1}
\bar{\sum}|\mathcal{M}{}^({}'{}^)|^2,
\end{equation}
where\
\begin{equation}
\eta = 1 + \frac{2\varepsilon_1}{M_p}\sin^2\frac{\theta}{2},\qquad \frac{\varepsilon_3}{\varepsilon_1} = \frac{1}{\eta},\qquad
\frac{\varepsilon_3'}{\varepsilon_1} = \frac{1}{\eta}\left(1 - \frac{M_\Delta^2-M_p^2}{2M_p \varepsilon_1}\right).
\end{equation}

The squared matrix elements can be presented as the products of current tensors:
\begin{equation}\label{eq:M2}
\bar{\sum} \left|\mathcal{M}\right|^2
= \frac{Z^2e^4}{(q^2)^2}\ 
L_{\nu\rho}(p_1, p_3)\ 
T_p^{\nu\rho}(p_2, p_4)
\end{equation}
and
\begin{equation}\label{eq:M2prime}
\bar{\sum} \left|\mathcal{M}'\right|^2
= \frac{Z^2e^4}{(q'{}^2)^2}\ 
L_{\nu\rho}(p_1, p_3')\ 
T_{p\to\Delta}^{\nu\rho}(p_2, p_4').
\end{equation}
The electron current tensor is
\begin{equation}
L^{\nu\rho}(p_1, p_3) =\bar{\sum}\,j^\nu(p_1, p_3)\,j^{\dagger\rho}(p_1, p_3).
\end{equation}
Proton $T_p$ and transition $T_{p\to \Delta}$ current tensors have the same
form. All of them are either well known or could be calculated straightforward.
We present appropriate formulas in the Appendix~A. 

The convolution in \eref{eq:M2} leads to the well known
Rosenbluth formula
\begin{equation}
\frac{\rmd\sigma}{\rmd\Omega} =
\frac{Z^2 \alpha^2\cos^2{\frac{\theta}{2}}}
{4\varepsilon_1^2\eta\sin^4{\frac{\theta}{2}}}\ \frac{\tau G_{\rm M}^2(q^2) + \epsilon
	G_{\rm E}^2(q^2)}{\epsilon(1+\tau)},
\end{equation}
where $\alpha =  e^2/4\pi$,
\begin{equation}\label{eq:tau_epsilon}
\tau = \frac{-q^2}{4M_p^2},\qquad \epsilon = \left(1 + 2(1+\tau)\tan^2\frac{\theta}{2}\right)^{-1}.
\end{equation}

Using \eref{eq:M2prime} for the process $ep\to e\Delta$ one can find
\begin{eqnarray}
    \frac{\rmd\sigma'}{\rmd\Omega} =&
\frac{Z^2 \alpha^2 \cos^2{\frac{\theta}{2}}} {4 \varepsilon_1^2 \eta
	\sin^4{\frac{\theta}{2}}}\frac{(M_p+M_\Delta)^2}{4M_p^2}\nonumber\\
    &\times\frac{\tau' \left( G_{\rm M}^{*2}(q'{}^2)+3 G_{\rm E}^{*2}(q'{}^2) + \epsilon' \frac{-q'{}^2}{M_\Delta^2}G_{\rm C}^{*2}(q'{}^2)\right)}{\epsilon'(1+\tau')},
\end{eqnarray}
where
\begin{equation}\label{eq:tau_epsilon_prime}
\tau' = \frac{-q'{}^2}{(M_\Delta+M_p)^2},\qquad
\epsilon' = \left(1 + 2\left(1+\frac{\nu^2}{-q'{}^2}\right)\tan^2\frac{\theta}{2}\right)^{-1},
\end{equation}
with $\nu = \varepsilon_1 - \varepsilon_3' = (M_\Delta^2 - M_p^2-q'{}^2)/(2M_p)$.

If we consider the cross sections ratio
\begin{eqnarray}
\frac{{\rmd\sigma'}/{\rmd\Omega}}
{{\rmd\sigma}/{\rmd\Omega}}=&
\frac{\epsilon(1+\tau)}{\epsilon'(1+\tau')}\,
\frac{(M_p+M_\Delta)^2}{4M_p^2}\nonumber\\
&\times\frac{
    \tau'\left(G_{\rm M}^{*2}(q'{}^2)+3G_{\rm E}^{*2}(q'{}^2) + \epsilon'
\frac{-q'{}^2}{M_\Delta^2}G_{\rm C}^{*2}(q'{}^2)\right)}
{ \tau G_{\rm M}^2(q^2) + \epsilon G_{\rm E}^2(q^2) },
\end{eqnarray}
and take into account the conditions of the VEPP-3 experiment
\cite{PhysRevLett.114.062005}
we will find that the cross section ratio is about $1$.  So for the estimate
\eref{eq:rough_estimate} of $\Delta$ contribution to real radiative corrections
for elastic $ep$-scattering we could write the following expression
\begin{equation}
\delta_\Delta \simeq
\frac{{\rmd\sigma'}/{\rmd\Omega}}
{{\rmd\sigma}/{\rmd\Omega}}\,
\frac{\Gamma_{\Delta\to p\gamma}}{\Gamma_\Delta} \simeq 0.5\%.
\end{equation}
 We used the branching ${\rm Br}({\Delta\to N\gamma}) = 0.55-0.65\%$ from PDG
 \cite{Agashe:2014kda} and the values of the transition form factors
 $G_{\rm M,E,C}^*$ derived from the parametrization \eref{eq:ZY} and the equations
 \eref{eq:JS_ZY_relation}. It is a very rough estimate, moreover the numerical
 value seems significant for the VEPP-3 experimental results where the TPE
 effect is of order $1\%$.  So in the following we will present a more
 accurate calculation.

\section{Proton bremsstrahlung with Delta(1232) in the intermediate state}

Hereafter we consider the process $e(p_1)\, p(p_2) \to e(p_3)\,p(p_4)\,
\gamma(k)$, which contributes to real photon radiative corrections. There are two
Feynman diagrams for the proton bremsstrahlung with $\Delta$ in the intermediate
state (see Figure~\ref{fig:M_Delta_diagrams}):
\begin{equation}\label{eq:iMdelta_dec}
\rmi\mathcal{M}_\Delta = \rmi \mathcal{M}_\Delta^{(1)} + \rmi \mathcal{M}_\Delta^{(2)},
\end{equation}
where
\begin{equation}\label{eq:iMdelta1}
\rmi\mathcal{M}_\Delta^{(1)} = \frac{\rmi Z^2e^3}{q_e^2}\, j_\nu(p_1, p_3)\, \epsilon_\mu^*(k)\ \frac{
	\bar{U}(p_4)\, \Delta^{\mu\nu}(t;k,q_e)\, U(p_2)}
{t^2 - M_\Delta^2 + \rmi\Gamma_\Delta M_\Delta},
\end{equation}
with
\begin{equation}
\Delta^{\mu\nu}(t;k,q_e) = \Gamma_{\Delta \to \gamma p}^{\mu\alpha}(t, k)\,
(\hat{t} + M_\Delta)\mathcal{P}_{\alpha\beta}(t)
\,\Gamma_{\gamma p \to \Delta}^{\nu\beta}(t, q_e),
\end{equation}
and
\begin{equation}\label{eq:iMdelta2}
    \rmi\mathcal{M}_\Delta^{(2)} = \frac{\rmi Z^2e^3}{q_e^2}\, j_\nu(p_1, p_3)\,
    \epsilon_\mu^*(k)\ 
    \frac{ \bar{U}(p_4)\, \Delta^{\nu\mu}(t';-q_e,-k)\, U(p_2)}
         {t'^2 - M_\Delta^2}
\end{equation}
where we use $t = p_1+p_2-p_3$ and $t' = p_2 - k$, the electron momentum
transfer is $q_e = p_1 - p_3$, and the $\Delta$ propagator contains
\cite{Zhou:2014xka}
\begin{equation}
    \mathcal{P}^{\alpha\beta}(t) = -g^{\alpha\beta} +
    \frac{\gamma^\alpha\gamma^\beta}{3} + 
    \frac{\hat{t}\gamma^\alpha t^\beta + t^\alpha \gamma^\beta \hat{t}}{3t^2}.
    \label{eq:Pab}
\end{equation}
We save the width $\Gamma_\Delta$ for the first term $\mathcal{M}_\Delta^{(1)}$
because there is the resonance region when $t^2$ is close to $M_\Delta^2$, and
this region can give the main contribution of $\Delta$ to real radiative
corrections.  In the second term $\mathcal{M}_\Delta^{(2)}$ the real photon
emission moves away the amplitude from the resonance, so the omitted width can
not sufficiently change the results.

We will use additional simplification leaving in results only the terms, which
have minimal powers of the photon energy $\omega$ and the difference
$M_\Delta - M_p$ assuming
\begin{equation}\label{eq:approx}
    \omega \ll M_p,\quad M_\Delta - M_p \ll M_p.
\end{equation}
The first limit is a part of the traditional soft photon approximation. This
approximation is more suitable to experiments with magnetic spectrometers to
study the electron-proton elastic scattering cross section, where energy
restrictions on the unobservable photon and proton are rather strict. In the
VEPP-3 experiment energy cuts are conservative, so the contribution of hard
photons can be sufficient. The second limit allows us to sufficiently
simplify the result of traces calculation in $|\mathcal{M}_\Delta|^2$ and in the
interference of $\mathcal{M}_\Delta$ with the amplitude of electron
bremsstrahlung. We do not modify the denominators of $\Delta$ propagator since
they define the resonance behavior of the amplitude $\mathcal{M}_\Delta$. As for
the numerators in the soft photon limit this approximation means expansion
in terms of the small ratio $(M_\Delta - M_p)/M_p$ and saving only the
leading terms.

\subsection{Delta(1232) contribution to elastic cross section measurement experiments}

The square of the matrix element $|\mathcal{M}_\Delta|^2$ leads to $C$-even
contribution, so it has no influence on the ratio $R$ of the elastic
$e^\pm p$-scattering cross sections in the leading order of electromagnetic coupling
constant. But, in principal, it could affect the results of the experiments to
measure the unpolarized $e^-p$-scattering cross section.

As we supposed above the leading contribution of $|\mathcal{M}_\Delta|^2$ to
bremsstrahlung differential cross section comes from
\begin{equation}
    \bar{\sum}
    \left|\mathcal{M}_\Delta^{(1)}\right|^2  =  \frac{Z^4e^6}{(q_e^2)^2}\
\frac{	L_{\nu\nu'}(p_1, p_3)\,H^{\nu\nu'}(t; k, q_e)}{(t^2-M_\Delta^2)^2 + \Gamma_\Delta^2 M_\Delta^2},
\end{equation}
where
\begin{eqnarray}
\fl H^{\nu\nu'}(t; k, q_e) =\frac{(-g_{\mu\mu'})}{2}{\rm Tr}\left[
(\hat{p}_4+M_p)
\Delta^{\mu\nu}(t; k, q_e)
(\hat{p}_2+M_p)
\gamma^0
\left[\Delta^{\mu'\nu'}(t; k, q_e)\right]^\dagger
\gamma^0
\right]
\end{eqnarray}
Calculation of the trace and its convolution with $-g_{\mu\mu'}$ is
straightforward but tedious even using the approximation (\ref{eq:approx}). Some
details are presented in Appendix B.

Integrating with respect to the final proton momentum $\bi{p}_4$ in the
special frame, where $t = p_1 + p_2 - p_3$ has no spatial components (i.e. $t^0 =W,\ \bi{t} = 0$, where $W$ is defined by $W^2 = (p_{1}+p_{2}-p_{3})^2$), we come to
\begin{eqnarray}
    \frac{\rmd \sigma_\Delta^{(1)}}{\rmd \Omega} =
    \frac{1}{(4\pi)^2}\frac{1}{4 M_p^2 \eta}\
    \int
    \frac{\varepsilon_3 \rmd\varepsilon_3}{\varepsilon_{3,el}}\frac{M_p}{W}
    \int
    \frac{\omega^2 \rmd\Omega_\gamma}{(2\pi)^3 2 \omega}
    \bar{\sum} \left|\mathcal{M}_\Delta^{(1)}\right|^2,
    \label{eq:dsigmabrem}
\end{eqnarray}
where we use the limit $\varepsilon_1, \varepsilon_3 \gg m$:
\begin{equation}
    \frac{W^2 - M_p^2}{2M_p\eta} = \varepsilon_{3,{\rm el}} - \varepsilon_3,\qquad
    \eta=1+\frac{2\varepsilon_1}{M_p}\sin^2\frac{\theta}{2},\qquad
    \varepsilon_{3,{\rm el}} = \frac{\varepsilon_1}{\eta},
\end{equation}
$\varepsilon_{3,{\rm el}}$ is the final electron energy in the elastic
scattering process; the photon energy in the special frame comes from the relation $W^2 = (p_{4} + k)^2$:
\begin{equation}
\omega = \frac{W^2 - M_p^2}{2W},
\end{equation}
and $\int \rmd\Omega_\gamma$ means the integration with respect to the photon
directions in that special frame. The integration with respect to
$\rmd\varepsilon_3$ and $\rmd\Omega_\gamma$ in (\ref{eq:dsigmabrem}) must be performed
taking into account the particular experimental cuts. For the
experiments with magnetic spectrometers (for example, the SLAC experiment
\cite{PhysRevD.49.5671}) we set the lower bound on the final
electron energy $ \varepsilon_{3,{\rm el}} - \Delta E < \varepsilon_3 <
\varepsilon_{3,{\rm el}}$ and integrate over the total solid angle of the
final photon directions. As for the VEPP-3 experiment
\cite{PhysRevLett.114.062005}, where the final electron and proton are detected in
coincidence, there are a lower bound for the final electron energy
$\varepsilon_{3,{\rm el}} - \Delta E$ and the final proton angles cuts on the
difference between the elastic and measured values ($\Delta
\theta_p$~and~$\Delta \phi_p$).

\begin{figure}
    \centering
    \includegraphics[]{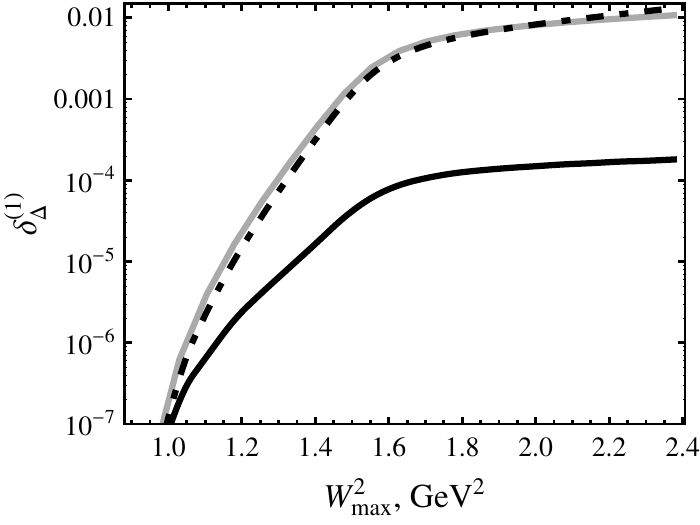}
    \caption{
        $\Delta(1232)$ contribution to real radiative corrections
        $\delta_\Delta^{(1)}$ for $E_{\rm beam} = 1.594\ {\rm GeV}$, $Q^2 =
        1.51\ {\rm GeV}^2$, i.e. for the point Run I, No.~1 at the VEPP-3
        experiment~\cite{PhysRevLett.114.062005}. Gray solid line represents the
        estimate \eref{eq:delta_1_approximate}; black dot-dashed line is
        numerical integration using the formula \eref{eq:dsigmabrem} with only
        $W_{\rm max}^2$ restriction; black solid line is numerical
        integration with  the proton emission angle cuts $\Delta \theta_p =
        \Delta \phi_p = 3\degree$ corresponded to the VEPP-3 experimental point
        Run I, No.~1.
    }
    \label{fig:dterm11}
\end{figure}

Using the approximation (\ref{eq:approx}) and the formulas from the Appendix B
we can find the contribution to real radiative corrections in the case of
spectrometric experiments on cross section measurements:
\begin{eqnarray}
    \delta_{\Delta}^{(1)} =& 
    \frac{ \rmd \sigma_\Delta^{(1)}/\rmd \Omega } { \rmd \sigma/\rmd \Omega
    }
    \approx 
    \frac{ \rmd \sigma'/\rmd \Omega } { \rmd \sigma/\rmd \Omega }\,
    \frac{\Gamma_{\Delta\to \gamma p}}{\Gamma_\Delta}
    \nonumber\\ 
    &\times\frac{1}{\pi} \int_0^{2M_p\eta \Delta E }
        \frac{\Gamma_\Delta M_\Delta }
            {(x-M_\Delta^2+M_p^2)^2 + \Gamma_\Delta^2 M_\Delta^2}
        \frac{x^3 \rmd x}{(M_\Delta^2-M_p^2)^3},
    \label{eq:delta_1_approximate}
\end{eqnarray}
where $x = W^2 - M_p^2$. The integral in the right hand side is almost obvious:
the square of $\Delta$~propagator yields the first multiplier; the powers of $x$
(or $\omega$) come from the photon phase space ($\omega^1$) and from the matrix
element, which starts with $\omega^1$ in the soft photon limit, so its square is
proportional to $\omega^2$; the integrand is proportional to
$\delta(x-M_\Delta^2+M_p^2)$ in the limit $\Gamma_\Delta \to 0$, therefore the
whole integral with the multiplier $1/\pi$ gives $1$ in that limit.

So, indeed, there is the term proportional to the cross section ratio multiplied
by the branching as we supposed in our rough estimate~(\ref{eq:rough_estimate}).
But it is also multiplied by the factor, which appears to be very small for
typical energy constraints in the experiments with magnetic spectrometers:
$W_{{\rm max}}^2  = M_p^2 + 2M_p\eta\,\Delta E < (M_p + m_\pi)^2$, i.e. below
the pion production threshold.

Numerical results for $\delta_\Delta^{(1)}$ are presented in
Figure~\ref{fig:dterm11}, where we show its dependence on the energy cut
$W_{\rm max}^2$. In the first case we do
not use any additional restrictions (elastic cross section measurements set-up).
One can see that the approximate formula~\eref{eq:delta_1_approximate} is in
quite good agreement with the full calculation of $|M_\Delta^{(1)}|^2$. 
The typical value $W_{{\rm max}}^2 = 1.12\ {\rm GeV}^2$ (from the
SLAC experiment \cite{PhysRevD.49.5671}) leads to a strong suppression of
$\Delta(1232)$ contribution.
In the second case we perform integration with the
final proton emission angle cuts which take place in the VEPP-3 experiment on
measurements cross section ratio $R$. Here we see that for the conservative
value $W_{{\rm max}}^2 \approx 1.6~{\rm GeV}^2$ in Run I, No. 1 the smallness of
the correction is primarily induced by the strict proton emission angle cuts.
Here and in the following the full calculation of traces and numerical Monte-Carlo
integration have been performed using FeynCalc
\cite{Mertig:1990an,Shtabovenko:2016sxi} and Wolfram Mathematica
\cite{Mathematica}.

\begin{figure}%
    \centering
    \includegraphics[]{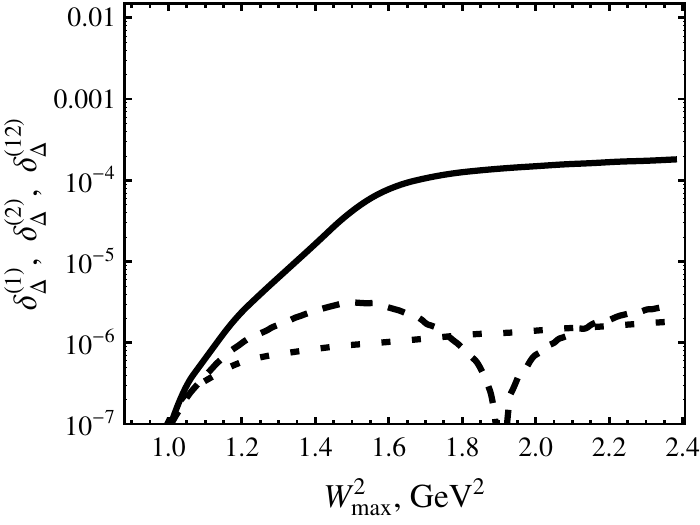}
    \caption{
        Various terms of $\Delta(1232)$ contribution to real radiative
        corrections: solid line shows $\delta_\Delta^{(1)}$, the
        contribution of $\left|\mathcal{M}_\Delta^{(1)}\right|^2$; dotted line
        is for $\delta_\Delta^{(2)}$, the contribution of
        $\left|\mathcal{M}_\Delta^{(2)}\right|^2$;
        dashed line is for absolute value $\left|\delta_\Delta^{(12)}\right|$ of
        the interference $2{\rm Re}\left[ \mathcal{M}_\Delta^{(1)}
        \mathcal{M}_\Delta^{(2)\dagger}\right]$
        (the interference changes the sign from positive to
        negative at about $W_{max}^2 \simeq 1.9\ {\rm GeV}^2$).
        Numerical integration is performed for $E_{\rm beam} = 1.594\ {\rm GeV}$, $Q^2 =
        1.51\ {\rm GeV}^2$ with the proton emission angle cuts $\Delta \theta_p =
        \Delta \phi_p = 3\degree$ corresponded to the VEPP-3 experimental point Run
        I, No.~1.
    }	
\label{fig:dterm}
\end{figure}
On the Figure~\ref{fig:dterm} one can find that our assumption about
$\mathcal{M}_\Delta^{(1)}$ dominance is in agreement with numerical results.
We compare the contributions of $|\mathcal{M}_\Delta^{(1)}|^2$,
$2{\rm Re}[ \mathcal{M}_\Delta^{(1)} \mathcal{M}_\Delta^{(2)\dagger}]$
and $|\mathcal{M}_\Delta^{(2)}|^2$. The second and the third
contributions are lower than the first one, as we supposed. It should be noted
that the regions $W^2 < M_\Delta^2$ and $W^2 > M_\Delta^2$
work in opposite directions for the interference, so in particular situations
the contribution can be suppressed and it will change the sign if
$W_{max}^2$ is sufficiently grater than $M_\Delta^2$.

\subsection{Delta(1232) contribution to real radiative corrections for the
VEPP-3 experiment.}

Here we investigate the $C$-odd interference of the proton bremsstrahlung with
$\Delta(1232)$ in the intermediate state and the electron bremsstrahlung. 
Assuming the approximation (\ref{eq:approx}) we decompose the electron
bremsstrahlung into traditional ``soft'' and ``hard'' parts
\begin{equation}
    \rmi\mathcal{M}_e = \rmi\mathcal M_e^{(s)} + \rmi\mathcal M_e^{(h)},
\end{equation}
\begin{equation}
    \rmi\mathcal M_e^{(s)} = -\frac{\rmi Ze^3}{q_p^2}\ j_\nu(p_1,p_3)\
    J_p^\nu(p_2, p_4)\
    \left[\frac{p_3^\mu}{(p_3 k)} - \frac{p_1^\mu}{(p_1 k)} \right]\,\epsilon_\mu^*(k),
\end{equation}
\begin{equation}
    \rmi \mathcal{M}_e^{(h)} = -\frac{\rmi Ze^3}{q_p^2}\ \bar{u}(p_3)
    \left(
        \frac{\gamma_\mu\hat{k}\gamma_\nu}{2(p_3 k)} +
        \frac{\gamma_\nu\hat{k}\gamma_\mu}{2(p_1k)}
    \right)
    u(p_1)\ J_p^\nu(p_2, p_4)\ \epsilon^{*\mu}(k),
\end{equation}
where $q_p = p_4 - p_2$ is the proton momentum transfer.

Our estimate for the interference is
\begin{equation}\label{eq:estimate}
    2\,{\rm Re}\left[ \bar {\sum} \mathcal{M}_e^\dagger \mathcal{M}_\Delta \right] \approx
    2\,{\rm Re} \left[\bar {\sum} \mathcal{M}_e^{(s)}{}^\dagger \mathcal{M}_\Delta^{(1)}\right].
\end{equation}
We can rewrite it as follows
\begin{equation}
    \bar{\sum} \mathcal M_e^{(s)}{}^\dagger \mathcal M_\Delta^{(1)} =  
    \frac{Z^3e^6}{q_e^2 q_p^2}
    \left[
        \frac{p_{3,\mu}}{(p_3 k)} - \frac{p_{1,\mu}}{(p_1 k)}
    \right]\,
    \frac{L_{\nu\nu'}(p_1, p_3)\,G^{\mu\nu\nu'}(t; k, q_e)}
        {t^2-M_\Delta^2 + \rmi \Gamma_\Delta M_\Delta},
\end{equation}
where
\begin{eqnarray}
G^{\mu\nu\nu'}(t; k, q_e) =\frac{1}{2}\Tr\left[
(\hat{p}_4+M_p)\,
\Delta^{\mu\nu}(t; k, q_e)\,
(\hat{p}_2+M_p)\,
\Gamma_{\gamma p \to p}^{\nu'}(-q_p)
\right].
\end{eqnarray}
Some details can be found in the Appendix C. Here we present only
the result within our approximation (\ref{eq:approx}):
\begin{eqnarray}
\fl 2\,{\rm Re}\left[\bar{\sum} \mathcal M_e^{(s)}{}^\dagger \mathcal M_\Delta^{(1)}\right] \approx&  \frac{Z^3e^6}{(q^2)^2}
\frac{2(W^2-M_\Delta^2)}{(W^2-M_\Delta^2)^2 + \Gamma_\Delta^2
    M_\Delta^2}\frac{2G_1(0)}{3}\frac{(M_\Delta+M_p)}{M_\Delta^2}\nonumber\\
    &\times\frac{2M_p(KP)}{P^2}\,\left(G_{\rm E}(q^2)
    {G}_{\rm M}^*(q^2) + \frac{-q^2}{4M_pM_\Delta }G_{\rm M}(q^2)
{G}_{\rm C}^*(q^2)\right)\nonumber\\
&\times\left[\frac{p_{3,\mu}}{(p_3 k)} - \frac{p_{1,\mu}}{(p_1 k)} \right]\
(-g_{\lambda\lambda'})\epsilon^{\lambda\tau\rho\mu}t_\tau k_\rho \,
\epsilon^{\lambda'\tau'\sigma\nu} t_{\tau'}(q_{e})_\sigma\,K_{\nu},
\end{eqnarray}
where $q_p^2\approx q_e^2 \approx q^2$, $K = p_1+p_3$, $P = p_2+p_4$.

The contribution to real radiative corrections has the following form 
\begin{equation}
\delta_{\Delta}^{({\rm int})} =
\frac{
    \rmd\sigma_\Delta^{({\rm int})}/\rmd\Omega
}
{
    \rmd\sigma/\rmd\Omega
},
\end{equation}
with 
\begin{equation}\label{eq:delta_Delta}
\frac{\rmd\sigma_\Delta^{({\rm int})}}{\rmd\Omega} = \frac{1}{(4\pi)^2}\frac{1}{4 M_p^2 \eta}\
\int
\frac{\varepsilon_3 \rmd\varepsilon_3}{\varepsilon_{3,el}}\frac{M_p}{W}
\int
\frac{\omega^2 \rmd\Omega_\gamma}{(2\pi)^3 2 \omega}\
\bar{\sum}\, 2{\rm Re}\left[\mathcal{M}_e^\dagger \mathcal{M}_\Delta\right],
\end{equation}
where the integration area is restricted by particular experimental cuts,
as it was explained for the similar formula \eref{eq:dsigmabrem}.

One can easily find that in the special frame ($t^0 = W,\ \bi{t} = 0$) the dependence
of the interference in the soft photon approximation on the photon emission
direction is determined by the factor
\begin{eqnarray}
    \fl 2\,{\rm Re}\left[\bar{\sum} \mathcal M_e^{(s)}{}^\dagger
    \mathcal M_\Delta^{(1)}\right] &\propto& 
        \left[\frac{p_{3,\mu}}{(p_3 k)} - \frac{p_{1,\mu}}{(p_1 k)} \right]\
    (-g_{\lambda\lambda'})\epsilon^{\lambda\tau\rho\mu}t_\tau k_\rho \,
    \epsilon^{\lambda'\tau'\sigma\nu} t_{\tau'}(q_{e})_\sigma\,K_{\nu}
    \nonumber\\
    &&=W^2\left(\left[\bi{k} \times \left(\frac{\bi{p}_{3}}{(p_3 k)} -
\frac{\bi{p}_{1}}{(p_1 k)}\right) \right]\cdot\left[\bi{q}_e \times
\bi{K}\right]\right),
\label{eq:int_spec_frame}
\end{eqnarray}
where all vectors are considered in that special frame. Then the integration
over the total solid angle yields to
\begin{equation}
\int \rmd\Omega_\gamma\ \frac{\bi{k}}{(p_1k)} \propto \bi{p}_1,\quad
\int \rmd\Omega_\gamma\ \frac{\bi{k}}{(p_3k)} \propto \bi{p}_3,
\end{equation}
so the first cross product in \eref{eq:int_spec_frame} and the whole
interference within the approximations (\ref{eq:approx}) and (\ref{eq:estimate})
yields zero for the experiments where all final photon directions in the special
frame are possible (as it takes place in the experiments with magnetic
spectrometers considered earlier).

\begin{table}
    	\caption{\label{table:delta_results}$\Delta(1232)$ contribution to real radiative corrections in the VEPP-3 experiment~\cite{PhysRevLett.114.062005}.}
    	\begin{indented}
    		\item[]
    		\lineup
        \begin{tabular}{@{}lcccc}
        	\br
            & Run I, No. 1 &  Run I, No. 2 & Run II, No. 1 &  Run II, No. 2 \\
            \mr
            $E_{\rm beam}\ ({\rm GeV})$ & 1.594 & 1.594 & 0.998 &   0.998\\
            $Q^2\ ({\rm GeV}^2)$ & 1.51 & 0.298 & 0.976 &   0.830\\
            $\Delta E/\varepsilon_{3,el}$ & 0.25 & 0.45 & 0.29 &   0.29\\
            $\Delta\theta_p,\ \Delta\phi_p$ & 3.0\degree$$ & 5.0\degree$$ & 3.0\degree$$ & 3.0\degree$$
            \\
            $\delta_{\Delta}^{(s,1)}, 10^{-5}$          &$ \m0.64\pm 0.03 $&$ 
              \m0.3\0\pm 0.1\0  $&$ \m2.96\pm 0.01$ &$  \m2.46\pm 0.01$\\
            $\delta_{\Delta}^{(h,1)}, 10^{-5}$          &$ -0.75\pm 0.01  $&$
            \m6.21\pm 0.05  $&$ -0.81\pm 0.01$ &$  -0.97\pm 0.01$\\
            $\delta_{\Delta}^{(s,2)}, 10^{-5}$          &$  \m1.26\pm 0.02 $&$
            -1.82\pm 0.03  $&$ \m1.20\pm 0.01$ &$  \m1.01\pm 0.01$\\
            $\delta_{\Delta}^{(h,2)}, 10^{-5}$          &$ -0.88\pm 0.01  $&$
            -1.58\pm 0.01  $&$ -0.49\pm 0.01$ &$  -0.55\pm 0.01$\\
            $\delta_{\Delta}^{({\rm int})}, 10^{-5}$   &$  \m0.32\pm 0.02  $&$  \m3.2\0\pm 0.1\0  $&$  \m2.86\pm 0.01$ &$  \m1.95\pm 0.01$\\
                \br
        \end{tabular}
        \end{indented}
\end{table}
But for the VEPP-3 experiment the integration over the total solid angle of the
photon emission directions in the special frame is consistent with the proton
angles cuts ($\Delta \theta_p\ {\rm and}\ \Delta\phi_p$) only within a certain
range of $\varepsilon_3$ (not too much different from the $\varepsilon_{3,el}$).
The actual area of integration with respect to the photon emission angles is
complex. So the result can only be computed numerically. In the Table
\ref{table:delta_results} we present the $\Delta(1232)$ contribution to real
radiative corrections for the VEPP-3 experiment: the ``soft'' and
``hard'' part of the interference with $\mathcal{M}_\Delta^{(1)}$ and
$\mathcal{M}_\Delta^{(2)}$; $\delta_\Delta^{(s,1)}$ comes from the contribution
of $\mathcal{M}_e^{(s)\dagger} \mathcal{M}_\Delta^{(1)}$ and so on; the values
are presented with  the estimates of Monte Carlo integration errors; the full
contribution $\delta_{\Delta}^{({\rm int})}$ was calculated independently on the
``soft'' and ``hard'' parts in order for an additional crosscheck, and within
the error it is in agreement with the sum of the partial contributions.
As one could expect, the results show a strong dependence on the experimental
conditions and cuts. We see that the soft photon approximation works better
for the Run II conditions, and for the Run I it gives the answer only in the
order of magnitude. Anyway the actual value of $\delta_{\Delta}^{({\rm int})} <
0.01\%$ ensures us that this contribution can not alter the results on the
$e^\pm p$ cross sections ratio where the TPE effect is about $1\%$.

\section{Conclusion}

Here we considered the contribution of $\Delta(1232)$ resonance to real
radiative corrections. It was shown that although the rough estimate gives the
significant value, the actual results are typically suppressed by strict energy
cuts or angular constraints. The effect is found to be negligible for past
experiments to measure unpolarized elastic scattering cross section as well as
for the recent experiment at the VEPP-3 storage ring to investigate the TPE
effects.

\section{Acknowledgments}

Work supported  by the Russian Foundation for Basic Research (grants 16-02-00888 and 15-02-02674).

\appendix
\section{Current tensors}
For the electron current tensor we have
\begin{eqnarray}
    L^{\nu\rho}(p_1, p_3) 
    &=\bar{\sum}\,j^\nu(p_1, p_3)j^{\dagger\rho}(p_1, p_3)
    \nonumber\\
    &=
    \frac{1}{2}{\Tr}\left[
        (\hat{p}_3 + m)
        \gamma^\nu
        (\hat{p}_1 + m)
        \gamma^\rho
    \right] \nonumber\\
    &= q^2 g^{\nu\rho} - q^\nu q^\rho + K^\nu K^\rho,
\end{eqnarray}
where $K = p_1 + p_3$, $q = p_1 - p_3$.

The proton current tensor is
\begin{eqnarray}
    \fl T_p^{\nu\rho}(p_2, p_4)=\bar{\sum}\,J_p^\nu(p_2,
    p_4)J_p^{\dagger\rho}(p_2, p_4)\nonumber\\ 
    =\frac{1}{2}\,{\rm Tr}\left[
    (\hat{p}_4 + M_p)\,
    \Gamma_{\gamma p \to p}^\nu(q)\,
    (\hat{p}_2 + M_p)\,
    \Gamma_{\gamma p \to p}^\rho(-q)\,
\right]\nonumber\\
=4M_p^2\Biggl[
\frac{-q^2}{4M_p^2}\ G_{\rm M}^2(q^2)\left( - g^{\nu\rho} + \frac{q^\nu q^\rho}{q^2} + \frac{P^\nu P^\rho}{P^2}
\right)+G_{\rm E}^2(q^2)\,\frac{P^\nu P^\rho}{P^2}\Biggr],
\end{eqnarray}
where $P = p_2 + p_4$, $q = p_4 - p_2$.

And the transition current tensor is
\begin{eqnarray}
\fl T_{p\to\Delta}^{\nu\rho}(p_2, p_4') = \bar{\sum}\,J_{p\to\Delta}^\nu(p_2, p_4')J_{p\to\Delta}^{\dagger\rho}(p_2, p_4')\nonumber\\
=\frac{1}{2}\, {\Tr}\left[
(\hat{p}_4' + M_\Delta)\mathcal{P}_{\alpha\beta}(p_4')\,
\Gamma_{\gamma p \to \Delta}^{\nu\beta}(p_4', q')\,
(\hat{p}_2 + M_p)\,
\Gamma_{\Delta\to\gamma p}^{\rho\alpha}(p_4', q')\,
\right]\nonumber\\
=\frac{(M_\Delta+M_p)^2}{4M_p^2} 
\, \left((M_\Delta - M_p)^2 - q'{}^2\right)\nonumber\\
\quad\times\Biggl[\left(G_{\rm M}^{*2}(q'{}^2) + 3G_{\rm E}^{*2}(q'{}^2)\right)
\left(-g^{\mu\nu} + 
\frac{q'{}^\mu q'{}^\nu}{q'{}^2}+
\frac{\tilde{P}^\mu
	\tilde{P}^\nu}{\tilde{P}^2}
\right)\nonumber\\ 
\qquad+ \frac{-q'{}^2}{M_\Delta^2}G_{\rm C}^{*2}(q'{}^2)\ 
\frac{\tilde{P}^\mu \tilde{P}^\nu}{\tilde{P}^2}
\Biggr],
\end{eqnarray}
where $q' = p_4' - p_2,$
\begin{equation}
\tilde{P}^\mu = P' - \frac{(P' \cdot q')}{q'{}^2}\,q'{}^\mu,\quad 
P' = p_2 + p_4',
\end{equation}
and we use the sum over $\Delta$-particle polarization states
\begin{equation}
\sum U_\alpha(t) \bar{U}_\beta(t) = (\hat{t} +
M_\Delta)\mathcal{P}_{\alpha\beta}(t),
\end{equation}
with $\mathcal{P}_{\alpha\beta}(t)$ defined in \eref{eq:Pab}.

For the elastic scattering process $ep\to ep$:\
\begin{eqnarray}
L_{\nu\rho}(p_1, p_3)\, T_p^{\nu\rho}(p_2,p_4) &=& \left((KP)^2 + q^2 P^2\right)
    \frac{\tau G_{\rm M}^2(q^2) + \epsilon G_{\rm E}^2(q^2)}{\epsilon(1+\tau)},
\end{eqnarray}
where\
\begin{equation}
\tau = \frac{-q^2}{4M_p^2},\quad \epsilon = \frac{(KP)^2 + q^2P^2}{(KP)^2 - K^2P^2 -2q^2P^2},
\end{equation}
and we used
\begin{equation}
(Kq) = 0,\quad (Pq) = 0.
\end{equation}

If we consider the case with ultrarelativistic electrons ($\varepsilon_1, \varepsilon_3 \gg m$, $q^2 \gg m^2$), we will have\
\begin{equation}
L_{\nu\rho}(p_1, p_3)\, T_p^{\nu\rho}(p_2,p_4) = 4M_p^2\left(4\varepsilon_1\varepsilon_3\cos^2\frac{\theta}{2}\right)
\frac{\tau G_{\rm M}^2(q^2) + \epsilon G_{\rm E}^2(q^2)}{\epsilon(1+\tau)}
\end{equation}
with $\tau$ and $\epsilon$ defined in (\ref{eq:tau_epsilon}).

As for the process $ep\to e\Delta$:
\begin{eqnarray}
    \fl L_{\nu\rho}(p_1, p_3')\, T_{p\to \Delta}^{\nu\rho}(p_2,p_4') =& \left((K'\tilde{P})^2 + q'{}^2 \tilde{P}^2\right)
    \frac{(M_\Delta+M_p)^2}{4M_p^2} 
    \nonumber \\
    &\times \frac{\tau'\left(G_{\rm M}^{*2}(q'{}^2) + 3G_{\rm E}^{*2}(q'{}^2) +
\epsilon' \frac{-q'{}^2}{M_\Delta^2} G_{\rm C}^{*2}(q'{}^2)\right)}{\epsilon'(1+\tau')},
\end{eqnarray}
where
\begin{equation}
\tau' = \frac{-q'{}^2}{(M_p+M_\Delta)^2},\quad \epsilon' = \frac{(K'\tilde{P})^2 + q'{}^2\tilde{P}^2}{(K'\tilde{P})^2 - K'{}^2\tilde{P}^2 -2q'{}^2\tilde{P}^2},
\end{equation}
and we used 
\begin{equation}
(K'q') = 0,\quad (\tilde{P}q') = 0,
\end{equation}
\begin{equation}
\tilde{P}^2 = \frac{P'{}^2 q'{}^2 - (P'q')^2}{q'{}^2} = \frac{\left((M_\Delta-M_p)^2-q'{}^2\right)\left((M_\Delta+M_p)^2-q'{}^2\right)}{-q'{}^2},
\end{equation}
where $K' = p_{1}-p_{3}'$.

Again in the case of ultrarelativistic electrons ($\varepsilon_1, \varepsilon_3' \gg m$, $q'{}^2 \gg m^2$) we have
\begin{eqnarray}
\fl L_{\nu\rho}(p_1, p_3')\, T_{p\to \Delta}^{\nu\rho}(p_2,p_4') =& 4M_p^2\left(4\varepsilon_1\varepsilon_3'\cos^2\frac{\theta}{2}\right)\frac{(M_\Delta+M_p)^2}{4M_p^2} 
\nonumber\\
&\times\frac{\tau'\left(G_{\rm M}^{*2}(q'{}^2) + 3G_{\rm E}^{*2}(q'{}^2) +
\epsilon' \frac{-q'{}^2}{M_\Delta^2} G_{\rm C}^{*2}(q'{}^2)\right)}{\epsilon'(1+\tau')},
\end{eqnarray}
with $\tau'$ and $\epsilon'$ defined in (\ref{eq:tau_epsilon_prime}).

\section{Approximation for  $\left|\mathcal{M}_\Delta^{(1)}\right|^2$}
To calculate the matrix element $\mathcal{M}_\Delta^{(1)}$ it is useful to
consider it in the special frame, where the 4-vector $t$ has no spatial
components
\(
    t = p_1 + p_2 - p_3,\ t = \{W, 0\}.
\)
We have in this special frame
\begin{equation}
    \eqalign{
        q_e =\{q_e^0,\bi{q}_e\},\qquad & p_2 = \{\mathscr{E}_2,-\bi{q}_e\},\\
        k = \{\omega, \bi{k}\},\qquad & p_4 = \{\mathscr{E}_4, -\bi{k}\},
    }
\end{equation}
where
\begin{equation}
	\eqalign{
	q_e^0 = \frac{W^2 - M_p^2 + q_e^2}{2W},\qquad&\mathscr{E}_2 = \frac{W^2 + M_p^2 - q_e^2}{2W},\\
	\omega = \frac{W^2 - M_p^2}{2W},\qquad&\mathscr{E}_4 = \frac{W^2+M_p^2}{2W},
}
\end{equation}
\begin{equation}
|\bi{q}_e | = \frac{\sqrt{(W-M_p)^2-q_e^2} \sqrt{(W+M_p)^2-q_e^2}}{2W}.
\end{equation}
The soft photon approximation means
\begin{equation}
W\to M_p,\qquad \mathscr{E}_4 \to M_p.
\end{equation}

One can easily check that in the special frame the numerator of the $\Delta$
propagator~(\ref{eq:Pab}) is equal to zero for time-like indexes:
\begin{eqnarray}\label{eq:Plimit}
\mathcal{P}^{0\beta}(t) = \mathcal{P}^{\alpha 0}(t) = 0.
\end{eqnarray}
For spatial indexes $a,b = 1,2,3$ we have (here and after we use Latin letters
for spatial components of 4-vectors and tensors):
\begin{eqnarray}
(\hat{t} + M_\Delta)\mathcal{P}^{ab}(t) \approx \frac{2 M_\Delta}{3}
\left(
\begin{array}{cc}
1 & 0\\
0 & 0
\end{array}
\right)\otimes
\left(2\delta^{ab} - i\epsilon^{abc}\boldsymbol\sigma^c\right),
\end{eqnarray}
where we have dropped the terms proportional to $W-M_\Delta$. Here we use the standard representation of the Dirac $\gamma$-matrices, the Pauli $\sigma$-matrices, and the spatial Levi-Civita tensor~$\epsilon^{abc}$.

Let us consider the vertex with the real photon emission in the special frame:
\begin{equation}
\Gamma_{\Delta \to \gamma p}^{0 a}(t,k) \approx
-\sqrt{\frac{2}{3}}\frac{W}{2M_\Delta^2}\left(
\begin{array}{cc}
0 & 1\\
1 & 0
\end{array}
\right)\otimes\left[
G_1(0)\, i\epsilon^{acd}\bi{k}^c\boldsymbol\sigma^d
-G_2(0)\,\bi{k}^a\right],
\end{equation}
and
\begin{eqnarray}
\fl \Gamma_{\Delta \to \gamma p}^{m a}(t,k) \approx&
-\sqrt{\frac{2}{3}}\frac{W}{2M_\Delta^2}\Biggl\{\left(
\begin{array}{cc}
0 & 1\\
1 & 0
\end{array}
\right)\otimes \left[
G_1(0)\,
i\epsilon^{amc}\omega\boldsymbol\sigma^c 
-G_2(0)\delta^{ma}\omega\right]\nonumber\\
&\quad-\left(
\begin{array}{cc}
1 & 0\\
0 & 1
\end{array}
\right)\otimes \left[
G_1(0)
\left(\delta^{ma}(\boldsymbol\sigma\bi{k})
-\boldsymbol\sigma^m \bi{k}^a\right)\right]
\Biggr\},
\end{eqnarray}
where we dropped the term with $G_3(0)$ because it is proportional to $\omega^2$. 

The vertex with the virtual photon absorption have the following form
\begin{eqnarray}
\fl \Gamma_{\gamma p \to \Delta}^{0 b}(t,q_e) =
-\sqrt{\frac{2}{3}}\frac{W}{2M_\Delta^2}\Biggl\{\left(
\begin{array}{cc}
0 & 1\\
1 & 0
\end{array}
\right)\otimes\left[
G_1(q_e^2)\,
i\epsilon^{bgf}\bi{q}_e^g \boldsymbol\sigma^f
+G_2(q_e^2)\bi{q}_e^b\right]\nonumber\\
\quad -\frac{G_3(q_e^2)}{M_\Delta}\left[ -\bi{q}_e^2
\left(
\begin{array}{cc}
-1 & 0\\
0 & 1
\end{array}
\right)\otimes\boldsymbol \sigma^b + 
q_e^0\bi{q}_e^b \left(
\begin{array}{cc}
0 & -1\\
1 & 0
\end{array}
\right) \right]
\biggr\}
\end{eqnarray}
and
\begin{eqnarray}
\fl \Gamma_{\gamma p \to \Delta}^{n b}(t,q_e) =
-\sqrt{\frac{2}{3}}\frac{W}{2M_\Delta^2}\Biggl\{\left(
\begin{array}{cc}
0 & 1\\
1 & 0
\end{array}
\right)\otimes\left[
G_1(q_e^2)\, i\epsilon^{bne}q_e^0\boldsymbol\sigma^e 
+G_2(q_e^2)\delta^{n b}q_e^0\right]\nonumber\\
\quad-\frac{G_3(q_e^2)}{M_\Delta}\left[
\left(q_e^2\delta^{nb} + \bi{q}_e^n\bi{q}_e^b\right)\left(
\begin{array}{cc}
0 & -1\\
1 & 0
\end{array}
\right) -
\bi{q}_e^n q_e^0\left(
\begin{array}{cc}
-1 & 0\\
0 & 1
\end{array}
\right)\otimes\boldsymbol \sigma^b
\right]\nonumber\\
\quad-\left(
\begin{array}{cc}
1 & 0\\
0 & 1
\end{array}
\right)\otimes \left[
G_1(0)
(\delta^{nb}(\boldsymbol\sigma\bi{q}_e)
-\boldsymbol\sigma^n \bi{q}_e^b)\right]
\Biggr\}.
\end{eqnarray}

In the soft photon limit the final proton bispinor has only top components
\begin{eqnarray}\label{eq:Ulimit}
U(p_4)\approx \left\{\sqrt{\mathscr{E}_4+ M_p}\ \varphi_4, 0 \right\},
\end{eqnarray}
while the bottom components contain to $\sqrt{\mathscr{E}_4 - M_p}\approx \sqrt{\omega^2/2M_p}$.

Taking into account the formulas (\ref{eq:Plimit})--(\ref{eq:Ulimit}) we can obtain the approximation\
\begin{eqnarray}\label{eq:Deltam0}
    \fl \Delta^{m0} \approx
    -\frac{G_1(0)}{9}\frac{W^2}{M_\Delta^3}\frac{\tilde{G}_{\rm C}(q_e^2)}{2}\left(
\begin{array}{cc}
0 & 1
\\
0 & 0
\end{array}
\right)\otimes
(2i\epsilon^{mlp}\bi{k}^l - \bi{k}^p
\boldsymbol\sigma^m + \delta^{mp}(\boldsymbol\sigma \bi{k})
)\bi{q}_e^p
\end{eqnarray}
and
\begin{eqnarray}\label{eq:Deltamn}
\fl \Delta^{mn} \approx
\frac{G_1(0)}{9}\frac{W^2}{M_\Delta^3}\Biggl\{-\frac{\tilde{G}_{\rm C}(q_e^2)}{2}\left(
\begin{array}{cc}
0 & 1 
\\
0 & 0
\end{array}
\right)\otimes(2i\epsilon^{mln}\bi{k}^l - \bi{k}^n
\boldsymbol\sigma^m + \delta^{mn}(\boldsymbol\sigma \bi{k})
)q_e^0
\nonumber\\
\fl\qquad\qquad+\frac{G_3(q_e^2)}{M_\Delta}
\left(
\begin{array}{cc}
0 & 1 
\\
0 & 0
\end{array}
\right)\otimes (2i\epsilon^{mlr}\bi{k}^l - \bi{k}^r
\boldsymbol\sigma^m + \delta^{mr}(\boldsymbol\sigma
\bi{k}))(\bi{q}_e^2\delta^{nr}-\bi{q}_e^n\bi{q}_e^r)
\nonumber\\
\fl\qquad\qquad+G_1(q_e^2)
\left(
\begin{array}{cc}
1 & 0
\\
0 & 0 
\end{array}
\right)\otimes\left[
\epsilon^{xml}\boldsymbol{\sigma}^l\epsilon^{xar}\bi{k}^r
(2\delta^{ab} - i\epsilon^{abc}\boldsymbol\sigma^c)
\epsilon^{x'nl'}\boldsymbol{\sigma}^{l'}\epsilon^{x'br'}\bi{q}_e^{r'}
\right]\Biggr\},
\end{eqnarray}
where we introduced
\begin{equation}
    \frac{\tilde{G}_{\rm C}(q_e^2)}{2} =  -\left(G_1(q_e^2) - G_2(q_e^2)\right)
+ G_3(q_e^2)\frac{q_e^0}{M_\Delta}.
\label{eq:GtildeC}
\end{equation}
Strictly speaking our approximation (\ref{eq:approx}) implies $W=M_{p}$, $q_{e}$
to be equal $q$ (the momentum transfer in the elastic scattering), no difference
between $M_\Delta$ and $M_p$ and some other relations. But since it is possible
to identify the presented terms in the full matrix element and trace
calculation results we do not perform all of these transformations here and in
the following section.

Taking into account that the integration with respect to all real photon directions leads to
\begin{equation}
\bi{k}^i\bi{k}^j \to \frac{ \omega^2}{3}\delta^{ij},
\end{equation}
we will write down the averaged value of 
\(\bar{H}^{\nu\nu'} = \int H^{\nu\nu'} \rmd\Omega_\gamma/4\pi\):
\begin{equation}
\eqalign{
\bar{H}^{00} =& \frac{G_1^2(0)}{9^2}\frac{W^4\omega^2}{M_\Delta^6}
    (\mathscr{E}_4+M_p)(\mathscr{E}_2-M_p)\,\tilde{G}_{\rm C}^2(q_e^2) \bi{q}_e^2,\nonumber\\
\bar{H}^{n0} =& \frac{G_1^2(0)}{9^2}\frac{W^4\omega^2}{M_\Delta^6}
    (\mathscr{E}_4+M_p)(\mathscr{E}_2-M_p)\,{\tilde{G}_{\rm C}^2(q_e^2)}\bi{q}_e^{n}q_e^0,\nonumber\\
\bar{H}^{nn'} =& \frac{G_1^2(0)}{9^2}\frac{W^4\omega^2}{M_\Delta^6}
    (\mathscr{E}_4+M_p)(\mathscr{E}_2-M_p)\Biggl\{\tilde{G}_{\rm C}^2(q_e^2)\frac{\bi{q}_e^n\bi{q}_e^{n'}}{\bi{q}_e^2} q_0^2\nonumber\\
&+
    M_\Delta^2\left(\tilde{G}_{\rm M}^2(q_e^2) +3 \tilde{G}_{\rm E}(q_e^2)\right)\left(\delta^{nn'}-\frac{\bi{q}_e^n\bi{q}_e^{n'}}{\bi{q}_e^2}\right)
\Biggr\},
}
\end{equation}
where it was  useful to introduce $\tilde{G}_{\rm M,E}$ in addition to $\tilde{G}_{\rm C}$~(\ref{eq:GtildeC}):
\begin{equation}
\eqalign{
    \frac{\tilde{G}_{\rm
    		M}(q_e^2) - \tilde{G}_{\rm E}(q_e^2)}{2} =  \frac{\mathscr{E}_2 + M_p}{M_\Delta}\,G_1(q_e^2),\\
    \tilde{G}_{\rm E}(q_e^2)  = -\frac{q_e^0}{M_\Delta}\left(G_1(q_e^2)-G_2(q_e^2)\right) + G_3(q_e^2)\frac{q_e^2}{M_\Delta^2}.
}
\end{equation}
These quantities can be reduced to $G_{\rm M,E,C}$ for $W = M_\Delta$
\begin{equation}
\tilde{G}_{\rm M,E,C}(q_e^2)\bigr|_{W=M_\Delta} = \frac{3(M_\Delta +
M_p)}{M_p}\,G_{\rm M,E,C}^*(q_e^2),
\end{equation}

The tensor $\bar{H}^{\nu\nu'}$ at the point $W = M_\Delta$ can be rewritten in terms of the transition current tensor $T_{p\to\Delta}$ and the partial width $\Gamma_{\Delta
	\to p\gamma}$:
\begin{equation}
\bar{H}^{\nu\nu'}\bigr|_{W=M_\Delta} \approx
\frac{64\pi\,\Gamma_{\Delta\to\gamma p}}{Z^2e^2}\frac{M_\Delta^5\,
	\omega^2}{(M_\Delta^2-M_p^2)^3}\ 
T_{p\to\Delta}^{\nu\nu'}(p_2, t)\bigr|_{W=M_\Delta},
\end{equation}
where
\begin{eqnarray}
\Gamma_{\Delta \to \gamma p}&= 
\frac{\bar{\sum} |\mathcal{M}_{\Delta \to \gamma p}|^2}{16\pi}\,\frac{M_\Delta^2-M_p^2}{M_\Delta^3}
= \frac{\left.-T^{\nu\nu}_{p\to \Delta}\right|_{W=M_\Delta, q^2 = 0}}{2} 
\frac{M_\Delta^2-M_p^2}{16\pi M_\Delta^3}\nonumber\\
&=\frac{Z^2e^2(M_\Delta^2-M_p^2)^3}{64\pi M_p^2 M_\Delta^3}\left[G_{\rm M}^{*2}(0)
+ 3G_{\rm E}^{*2}(0)\right]\nonumber\\
&\approx\frac{Z^2e^2(M_\Delta^2-M_p^2)^3}{144\pi
	M_\Delta^3}G_1^2(0).
\end{eqnarray}

Finally, we have the following expression for the differential cross section
(\ref{eq:dsigmabrem}):
\begin{eqnarray}\label{eq:delta_Delta_approx}
\frac{\rmd\sigma_\Delta^{(1)}}{\rmd\Omega} \approx& \frac{\rmd\sigma'}{\rmd\Omega}\ \frac{\Gamma_{\Delta\to\gamma p}}{\Gamma_\Delta}\nonumber\\
&\times\frac{1}{\pi}\int_{0}^{2M_p\eta \Delta E}
\left[\frac{\Gamma_\Delta M_\Delta }{(x+M_p^2-M_\Delta^2)^2 + \Gamma_\Delta^2 M_\Delta^2}\right] 
\frac{x^3\,\rmd x}{(M_\Delta^2-M_p^2)^3},
\end{eqnarray}
where we used our approximation (\ref{eq:approx}):
\begin{equation}
\eqalign{
x = W^2 - M_p^2,\qquad&  \rmd x \approx 2M_\Delta \rmd W,\\
W\rmd W = -M_p\eta\, \rmd\varepsilon_3,\qquad&\omega \approx \frac{x}{2M_\Delta},
}
\end{equation}
and
\begin{equation}
\frac{\rmd\sigma'}{\rmd\Omega} = \frac{1}{(4\pi)^2}\frac{1}{4 M_p^2 \eta}\
\frac{\varepsilon_3}{\eta\varepsilon_{3,el}}
\frac{Z^2e^2}{(q_e^2)^2}\
L_{\nu\nu'}(p_1, p_3)T_{p\to\Delta}^{\nu\nu'}(p_2, t)\biggr|_{W=M_\Delta}.
\end{equation}

\section{Approximation for the interference $\mathcal{M}_e^{(s)\dagger} \mathcal{M}_\Delta^{(1)}$}
Here we consider the tensor, which appears in the interference:
\begin{equation}
G^{\mu\nu\nu'}(t; k, q_e) =\frac{1}{2}\,{\Tr}\left[
(\hat{p}_4+M_p)\,
\Delta^{\mu\nu}(t; k, q_e)\,
(\hat{p}_2+M_p)\,
\Gamma_{\gamma p \to p}^{\nu'}(-q_p)
\right].
\end{equation}
Here for real protons one can make the substitution
\begin{equation}
\Gamma_{\gamma p \to p}^\nu(-q_p) = 
    2M_p(G_{\rm E}(q_p^2) - G_{\rm M}(q_p^2)) \frac{P^\nu}{P^2} +
G_{\rm M}(q_p^2) \gamma^\nu ,
\end{equation}
where $P = p_2 + p_4$.
Therefore it is possible to decompose the tensor $G$:
\begin{eqnarray}
    G^{\mu\nu\nu'}(t; k, q_e) = \frac{2M_pG_{\rm E}(q_p^2)}{P^2}\, P^{\nu'} G_1^{\mu\nu} + 
G_{\rm M}(q_p^2)\, G_2^{\mu\nu\nu'}.
\end{eqnarray} 

A straightforward calculation with approximate values of $\Delta^{m\nu}$ from
(\ref{eq:Deltam0}) and (\ref{eq:Deltamn}) leads to the following values in the
special frame:
\begin{equation}
    \eqalign{
        G_1^{m0} &= 0,\\
        G_1^{mn} &= \frac{G_1(0)}{9}\frac{W^2}{M_\Delta^3}\left(\mathscr{E}_4 + M_p\right)
((\bi{k}\bi{q}_e)\delta^{mn}- \bi{q}_e^m\bi{k}^n)\,
M_\Delta \tilde{G}_{\rm M}(q_e^2).
}
\end{equation}
It is worth to note that these tensors appear in convolution with the symmetric
tensor $L_{\nu\nu'}$.  The  symmetrized values for the second tensor
($\tilde{G}_2^{m\nu\nu'} = (G_2^{m\nu\nu'}+G_2^{m\nu'\nu})/2$):
\begin{equation}
\eqalign{
    \tilde{G}_2^{m00} =& 0,\\
    \tilde{G}_2^{m0n} =&
\frac{G_1(0)}{9}\frac{W^2}{M_\Delta^3}\left(\mathscr{E}_4 + M_p\right)
\frac{
	((\bi{k}\bi{q}_e)\delta^{mn}-\bi{q}_e^m\bi{k}^{n})}
{2}\\
&\times\left[
\frac{\left(\mathscr{E}_2 - M_p\right)}{2}\tilde{G}_{\rm C}(q_e^2)
+ \left(1-\frac{2M_p(\mathscr{E}_2+\mathscr{E}_4)}{P^2}\right) M_\Delta
\tilde{G}_{\rm M}(q_e^2)
\right],\\
\tilde{G}_2^{mnn'} =& \frac{G_1(0)}{9}\frac{W^2}{M_\Delta^3}
\left(\mathscr{E}_4 + M_p\right)\\
&
\times\left(
\bi{k}^{n'} \bi{q}_e^m \bi{q}_e^n
-\bi{q}_e^n (\bi{k} \bi{q}_e) \delta^{mn'}
+\bi{k}^{n} \bi{q}_e^m \bi{q}_e^{n'}
-\bi{q}_e^{n'} (\bi{k}\bi{q}_e) \delta^{mn}
\right)\\
&\times\left[G_1(q_e^2)-
G_3(q_e^2)\frac{\left(\mathscr{E}_2 - M_p\right) }{2M_\Delta} -
\frac{M_pM_\Delta}{P^2} \tilde{G}_{\rm M}(q_e^2)\right],
}
\end{equation}
where we used $\bi{P}\approx -\bi{q}_e$ in the soft photon limit.

Finally, using (\ref{eq:approx}) we can find the approximate result
\begin{eqnarray}
G^{\mu\nu\nu'}(t; k, q_e) \approx&
\frac{2G_1(0)}{3}\frac{(M_\Delta+M_p)}{M_\Delta^3}\frac{2M_p}{P^2}\nonumber\\
&\times \left(M_\Delta
    G_{\rm E}(q_p^2)
    {G}_{\rm M}^*(q_e^2) + \frac{-q_e^2}{4M_p}G_{\rm M}(q_p^2)
{G}_{\rm C}^*(q_e^2)\right)\nonumber\\
&\times P^{\nu'}(-g_{\lambda\lambda'})\epsilon^{\lambda\tau\rho\mu}t_\tau k_\rho \,
\epsilon^{\lambda'\tau'\sigma\nu} t_{\tau'}(q_{e})_\sigma\ .
\end{eqnarray}

\section*{References}
\bibliography{delta_rc_article}
\bibliographystyle{h-physrev}
\end{document}